\documentclass[aps,prb,preprint,superscriptaddress,showpacs,preprintnumbers,amsmath,amssymb]{revtex4-1}

\usepackage{graphicx}
\usepackage{epstopdf}
\usepackage{comment}
\usepackage{amsmath}
\usepackage{verbatim}

\newcommand{\1}[1]{#1^{(1)}}
\newcommand{\2}[1]{#1^{(2)}}
\newcommand{\DifrOp}[1]{\left[\left[#1\right]\right]}

\begin{document}

\preprint{}

\title{Dynamics of charged domain walls in ferroelectrics}

\author{M. Y. Gureev}
\email{maxim.gureev@epfl.ch}
\affiliation{Ceramics Laboratory, Swiss Federal Institute of Technology (EPFL), CH-1015 Lausanne, Switzerland}

\author{P. Mokr\'{y}}
\affiliation{Faculty of Mechatronics, Informatics and Interdisciplinary Studies, Technical University of Liberec, CZ-46117 Liberec, Czech Republic}
\author{A. K. Tagantsev}
\affiliation{Ceramics Laboratory, Swiss Federal Institute of Technology (EPFL), CH-1015 Lausanne, Switzerland}
\author{N. Setter}
\affiliation{Ceramics Laboratory, Swiss Federal Institute of Technology (EPFL), CH-1015 Lausanne, Switzerland}

\date{\today}

\begin{abstract}

The interaction of electric field with charged domain walls in ferroelectrics is theoretically addressed.
A general expression for the force acting per unit area of a charged domain wall carrying free charge is derived.
It is shown that, in proper ferroelectrics, the free charge carried by the wall is dependent on the size of the adjacent domains.
As a result, it was found that the mobility of such domain wall (with respect to the applied field) is sensitive to the parameters of the domain pattern containing this wall.
The problem of the force acting on a planar charged $180^\circ$ domain wall normal to the polarization direction in a periodic domain pattern in a proper ferroelectric is analytically solved in terms of Landau theory.
It is shown that, in small applied fields (in the linear regime), the forces acting on walls in such pattern increase with decreasing the wall spacing, the direction of the forces coinciding with those for the case of the corresponding neutral walls.
At the same time, for large enough wall spacings and large enough fields, these forces can be of the opposite sign.
It is shown that the domain pattern considered is unstable in a defect-free ferroelectric.
The poling of a crystal containing such pattern,  stabilized by the pinning pressure, is also considered.
It is shown that, except for a special situation, the presence of charge domain walls can make poling more difficult.
It is demonstrated that the results obtained are also applicable to zig-zag walls under the condition that the zig-zag amplitude is much smaller than the domain wall spacing.
\end{abstract}

\maketitle

\section{Introduction}\label{Introduction}

Typically, at domains walls in ferroelectrics, the normal component of the electrical displacement is conserved with high accuracy.
This is the so-called condition of electrical compatibility\cite{Tagantsev_book}.
When a ferroelectric behaves as a perfect insulator, any appreciable violation of this condition will imply the appearance of  bound charge at the wall and a macroscopic electric field in the adjacent domains.
Such a field, at best, can  strongly increase the energy of the system, but very often this field is expected to be strong enough to fully suppress the ferroelectricity in the sample.
At the same time, the walls with an essential violation of the electrical compatibility - the so-called charged domain walls - have been observed in different materials such as lead titanate PbTiO$_3$\cite{Fesenko73,Fesenko85,Surowiak},
Pb[Zr$_x$Ti$_{1-x}]$O$_3$ (PZT)\cite{Jia}, lead germanate Pb$_5$Ge$_3$O$_{11}$ (PGO)\cite{Shur93,Shur}, and lithium niobate \cite{kugel,nakamura}.
It is believed that such walls can exist due to the compensation of its bound  charge by free electron\cite{Fesenko73,Fesenko85,Surowiak} or ion charges \cite{kugel,nakamura}.
Here, the situation with the electron compensation is of special interest.
Specifically, as was recognized many yeas ago, the concentration of free carriers in properly compensated  domain walls can readily achieve the metallic level \cite{Vul,Ivanchik}.
Thus, one can treat such a wall as a highly conductive interface.
In principle, one can tune its conductivity by controlling the wall orientation \cite{Morozovska} and change its position in the crystal by the application of a dc electric field.
In this context,  mobility of charged domain walls is an important issue.
Additionally, this issue deserves attention in view of poling samples containing charged domains walls.

The dynamics of a charged domain wall has not been completely understood and theoretical results are limited.
Here, one can mention only a classical paper by Landauer \cite{Landauer} and a recent paper by Mokry et al.\cite{Mokry}
Landauer indicated that the compensation of the bound charge on a wall will lead to a reduction of the pressure acting  on it in the presence of a dc electric field, whereas Mokry et al. qualitatively describled this effect in terms of Landau theory.
A closer analysis of the problem, however, shows that it misses a more involved theoretical treatment.
For example, a straightforward application of the results by  Mokry et al. \cite{Mokry} to the case of a "zig-zag" charged wall leads to a paradox.

The goal of this paper is to revisit the  problem of mobility of charged domain walls.
The  paper is organized as follows.

In Sect. \ref{sec:Generalized} we discuss and solve the aforementioned paradox.
Here, we obtain the general expression for the local force acting on the domain wall.
It is shown that not only the normal component of the force (treated in Ref.\cite{Mokry}), but also the tangential one should be considered.
In Sect. \ref{Force} we considered the force acting of a planar charged $180^\circ$ domain wall normal to the polarization direction in a periodic domain pattern in a proper ferroelectric.
We have shown that, for a fixed applied field, the force acting on the domain wall is a function of the structure period.
The results obtained are applied to the analysis of poling of such a structure.
Section \ref{Discussion} is devoted to a discussion of the implications of the results obtained to the experimental situation and available experimental data.
In Appendix, supporting calculations for Sect. \ref{Force} are presented.

\section{Local force density on a ferroelectric domain wall}
\label{sec:Generalized}

\begin{figure}
    \includegraphics[width=0.8\textwidth]{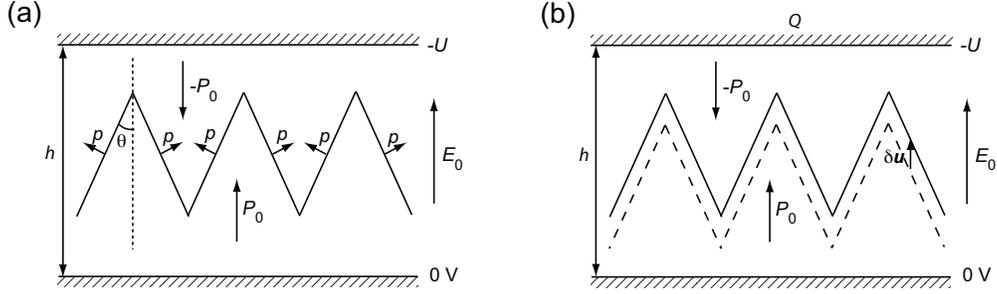}
    \caption{Zig-zag wall in a parallel plate capacitor. The pressure acting on each segment leads to non-zero force acting on the wall (a).
    During the virtual displacement of the "zig-zag" domain wall the charge $Q$ on the top electrode remains unchanged.
    Thus the force on the wall, calculated directly with the principle of virtual displacement, is zero (b).}
    \label{zig_zag_Fig}
\end{figure}

In this section we will consider the problem of the force acting on a charged domain wall placed in an electric field.
Originally, this problem was addressed by Mokry et al. \cite{Mokry}, who obtained a relation linking the pressure acting on the field (the surface density of the normal component of the force) with values of polarization and electric field in the adjacent domains and surface density of the free charge carried by the walls.
However, one finds a situation where a straightforward application of this result leads to a paradox.
This can be demonstrated for the case of  the force acting on charged "zig-zag" walls, which are typically observed in experiments \cite{Shur,Fesenko73,Fesenko85,Surowiak}.
Consider a fully compensated (the total (bound+free) charge on the wall equal to zero) zig-zag wall in a parallel plate capacitor of a thickness $h$, shown in Fig. \ref{zig_zag_Fig}.
Let us do this in hard ferroelectric approximation where the electrical displacement inside the domains is presented as a sum of the spontaneous polarization and the linear dielectric response to the electric field with permittivity $\varepsilon_\mathrm{f}$.
The bottom electrode is grounded and the top electrode is connected to a source of constant voltage $-U$.
In this case,  the pressure on each flat element of the wall was obtained by Mokry et al.\cite{Mokry} as:
\begin{equation} \label{pressure_full_comp}
p=2P_0E_0\cos^2 \theta,
\end{equation}
where $P_0$ is the spontaneous polarization, $E_0=U/h$ is the applied field, $\theta$ is the angle between the wall and the applied electric field.
The directions of the  corresponding forces are shown in Fig.\ref{zig_zag_Fig}a.
As is clear from this figure, such expression for the pressure leads to a nonzero total force acting on the zig-zag structure.

On the other hand, we can obtain this force directly from the principle of virtual displacement.
This method leads to a zero force acting on the fully compensated zig-zag domain wall.

 Indeed, according to this principle, the work $\delta W$ done  by the external force $F$
is equal to the variation of the proper thermodynamic function $\delta G$ during virtual displacement $\delta u$: $\delta W=F\delta u=\delta G$.
For the system considered, in the hard ferroelectric approximation, this thermodynamic function can be defined as \cite{Mokry}:
\begin{equation}\label{G_full_comp}
G=\int_V\frac{1}{2}\varepsilon_{\mathrm{f}}E^2dV + UQ,
\end{equation}
where $Q$ is the charge on the top electrode, $E$ is the electric field.
In view of full charge compensation, the electric field inside the ferroelectric is equal to $U/h$ and does not change with virtual displacement.
The virtual displacement of the "zig-zag" wall will not affect the charges on the electrodes (see Fig. \ref{zig_zag_Fig}(b)),
thus the thermodynamic function $G$ will not change with the virtual displacement.
This implies that the force acting on the wall is zero.

The above paradox can be resolved if we take into account the fact that the local force is not necessarily perpendicular to the domain wall.
Thus to calculate the resultant force, acting on an element of the domain wall, it is not sufficient to know the pressure.
The component of the force parallel to the domain wall should also be found.
As we will see later, in the case of the fully compensated wall, the consideration of all components of the force density will give the resultant force equal to zero.

Let us obtain the general expression for the force density on the domain wall.
We will start from scratch using the principle of virtual displacement.
We can not use the formula for the generalized stress tensor from textbook by Landau and Lifshitz \cite{Landau},
as was done in Ref. \cite{Mokry}.
In fact, no derivation of this formula is available.
In addition, one can show that it leads to results which are in contradiction with the results obtained with the principle of virtual displacement.

In our analysis we will use the same approximations as those used by Mokry et al.\cite{Mokry}.
First, the internal structure of domain wall is neglected.
Second, the free charge is moving together with the domain wall.
Third, we will not consider the forces arising due to an imbalance of the elastic energy, which is a suitable approximation for non-ferroelastic domain walls.

In order to obtain a general formula for the local force density
on domain wall, we follow the principle of virtual displacements,
as applied to the domain wall of arbitrary shape.
We will consider the problem at fixed potentials on the conductors $\varphi_E^{(i)}$.
We will use the thermodynamic function $G$ of the general form:
\begin{equation}\label{G}
    G = \int_V \left[\Phi(P) + \frac 12\varepsilon_\mathrm{b} E^2\right]\, dV - \sum_i\varphi_E^{(i)}\, Q_E^{(i)},
\end{equation}
where the first term represents the part of the free energy, $\Phi(P)$, associated with the ferroelectric part of polarization $P$ and the electric field energy integrated over the volume $V$ of the ferroelectric ($E$ is the electric field);
the second term represents the subtracted work of the electric sources ($Q_E^{(i)}$ are the charges on the conductors).
Here, $\varepsilon_\mathrm{b}$ is the background permittivity, which  was for simplicity taken as isotropic.
Thus, in our notation, the vector of electric displacement reads:
\begin{equation}\label{D}
D_i=\varepsilon_\mathrm{b}E_i+P_i.
\end{equation}
More or less straightforward analysis in terms of virtual displacement, given in Appendix \ref{appendix_force}, leads to the following expression for the force acting on a unit area of the the domain wall:

\begin{equation}
    \label{force_simplified}
    f_k =
    \DifrOp{\Phi(P)} n_k
        - \widehat{E}_i \DifrOp{P_i} n_k +
       \widehat{E}_k \sigma_\mathrm{f},
\end{equation}
where $\DifrOp{Z}=\2{Z}-\1{Z}$ denotes the jump of the quantity $Z$ on the domain wall, $ \widehat{E}=(\1{E}+\2{E})/2$, and $n_k$ is a unit vector, normal to the domain wall, directed from domain (1) to domain (2).
The upper suffix is used to designate the domains.

First two terms in Eq. \eqref{force_simplified} are the same as for the neutral domain wall.
The third term is the Coulomb force acting on the free charge on the domain wall.

We can rewrite Eq. \eqref{force_simplified} as the sum of the components perpendicular and parallel to the domain wall as:
\begin{equation}
    \label{force_components}
    f_k = \left\{
    \DifrOp{\Phi(P)}
        - \widehat{E}_i\left(\DifrOp{P_i} -
        \sigma_f n_i\right)
        \right\} n_k +
       \left\{
       \widehat{E}_k- \widehat{E}_j n_j \, n_k
       \right\}\sigma_\mathrm{f}.
\end{equation}
The first term in the curly brackets is the pressure acting on the domain wall,
which is the same as that found before by Mokry et al. \cite{Mokry}:
\begin{equation}
    \label{pressure}
 p=
    \DifrOp{\Phi(P)}
        - \widehat{E}_i\left(\DifrOp{P_i} -
        \sigma_f n_i\right).
\end{equation}
In this particular case the results, obtained with the principle of virtual displacement and
with the generalized stress tensor from the Landau book \cite{Landau} are the same\cite{Landau_comment}.

The second term in Eq. \eqref{force_components} represents
 the force parallel to the domain wall
 \begin{equation}
    \label{force_parallel_local}
    f_{||} =\widehat{E}_{||} \sigma_\mathrm{f},
\end{equation}
where $\widehat{E}_{||k}= \widehat{E}_k- \widehat{E}_j n_j \, n_k$ is the component of the average electric field parallel to the domain wall.
The force $f_{||}$ is the tangential component of the Coulomb force acting on the free charges on the wall.

Let us use the result obtained to find the force acting on a fully compensated zig-zag domain wall, the situation where we have come across a paradox.
As before, we consider the problem in the hard ferroelectric approximation.
We consider the model of a thin capacitor, with the thickness $h$, much smaller than the capacitor width.
It is the same model as used to obtain the well-known formula for the capacitance of the thin parallel-plate capacitor.
In this model, to exclude the edge effects and to consider a uniform electric field inside the capacitor, we should deal with the force per unit area of the electrode in the limit of infinitely wide capacitor instead of the total force acting on the wall.

The free charge density on the completely compensated "head-to-head" domain wall, that is equal to the jump of the normal component of the spontaneous polarization, reads:
  \begin{equation} \label{free_charge_compensated}
\sigma_\mathrm{f} =-2P_0\sin\theta.
\end{equation}
Thus, from Eq. \eqref{force_components} one finds the force acting on a flat segment of the domain wall as
\begin{equation} \label{force_compensated}
\begin{split}
F_k =&\{2P_0E_0\cos^2\theta\, n_k-2P_0\sin \theta(E_{0k}-E_0\sin\theta\, n_k)\}S=\\
=&2P_0E_0S \left( n_k-\sin\theta\frac{E_{0k}}{E_0} \right),
\end{split}
\end{equation}
where $S$ is the area of the wall segment.
This force is perpendicular to the direction of the electric field, i.e. it is parallel to the capacitor electrodes.
Indeed, one can check that
\begin{equation} \label{force_0}
F_k E_{0k}=2P_0E_0S \left( n_kE_{0k}-E_0\sin\theta\right)=0.
\end{equation}
The second term in brackets in Eq. \eqref{force_compensated} is parallel to the electric field, thus it is not contribute to the component of the force parallel to the capacitor electrodes, and  the force can be found as 
\begin{equation} \label{force_vector}
F_k=2P_0E_0Sd_k,
\end{equation}
where $d_k$ is the projection of $n_k$ on the plane of the electrodes.

\begin{figure}[h]
\begin{center}
\includegraphics[width=0.8\textwidth]{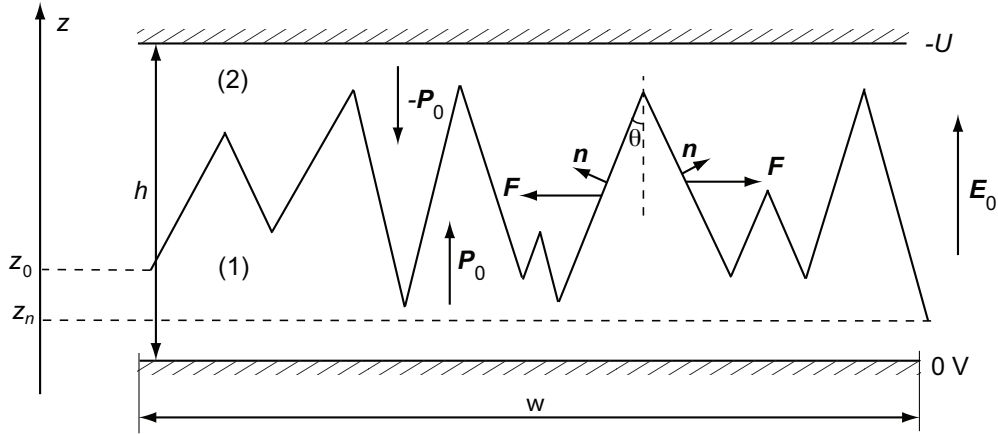}
\caption{Schematic of the zig-zag domain wall in a thin ferroelectric capacitor.
The polarizations inside the domains are perpendicular to the electrodes.
The bottom electrode is grounded and the constant potential, $-U$, is applied to the top electrode.
The electric field $E_0$ is uniform.
The resultant force acting on the zig-zag wall depends only on the coordinates $z_0$ and $z_n$ of the side points.
This force represents the edge-effect, which can be neglected in the thin capacitor model.
Neglecting the edge-effects, the resultant force acting on the domain wall is zero. }
\label{zig-zag}
\end{center}
\end{figure}

 Thus the absolute value of the force can be found from Eq. \eqref{force_vector} as:
\begin{equation} \label{force_absolute}
|F|=2P_0E_0S\cos\theta.
\end{equation}
This force is acting in the direction of a vector $d_k$, i.e. from domain 1 to domain 2. 
Thus forces acting on the neighbor segments are oppositely directed (see Fig. \ref{zig-zag}).
Thus the total force acting on the zig-zag wall containing $n$ segments can be found as the sum the the forces acting on each segment
\begin{equation}
F= 2P_0E_0\sum_{i=1}^{n}{(-1)^{i}S_{i}}\cos\theta_i=2P_0E_0l\sum_{i=0}^{n}{(z_i-z_{i-1})}=2P_0E_0l(z_n-z_0),
\end{equation}
where the $z$-axis is directed perpendicular to the electrodes, $z_i$ is the coordinate of the zig-zag vertex, $l$ is the capacitor width in the direction, perpendicular to the figure plane (Fig. \ref{zig-zag}).
The force obtained depends only on the coordinates of the vertices on the edges of the capacitor.
Considering the density of this force per unit area of the electrode we get
\begin{equation}
\frac{F}{S_\mathrm{el}}=\frac{2P_0E_0l(z_n-z_0)}{lw}\leq\frac{2P_0E_0h}{w},
\end{equation}
where $w$ is the width of the capacitor (see Fig. \ref{zig-zag}).
In the thin capacitor model $h/w\to 0$, thus the force density
 vanishes.
Thus, the total force acting on the wall is zero, and the above paradox is resolved.

\begin{figure}[h]
\begin{center}
\includegraphics[width=0.35\textwidth]{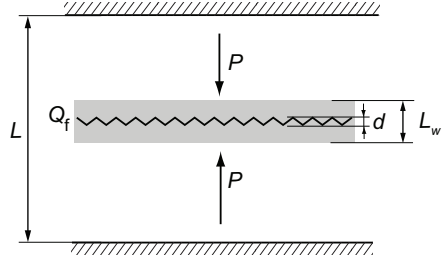}
\caption{Schematic of zig-zag domain walls in a ferroelectric capacitor.
The polarizations inside domains are perpendicular to electrodes.
The zig-zag amplitude $d$ is much smaller than the domain size.
This wall can be considered together with a layer around it of thickness $L_\mathrm{w}$ ($d\ll L_\mathrm{w}\ll L$), which is shown with the dark region on the figure.
The polarizations inside the bulk of the domains, the free electric charges on the walls, and the
force acting on the walls are the same for the flat and zig-zag walls.
}
\label{small_zig_zag}
\end{center}
\end{figure}

Concluding this Section we would like to discuss the applicability of Eq. \eqref{force_simplified} to the case of a partially compensated  zig-zag wall.
To calculate the force acting on such wall, one should know the electric field and ferroelectric part of polarization on both sides of the wall and apply Eq. \eqref{force_simplified} locally.
In general, this is a tough mathematical problem.
However, a simple answer can be obtained in the case where the  zig-zag amplitude $d$ is much smaller than the domain size $L$ (see Fig. \ref{small_zig_zag}).
We can consider such a zig-zag wall  together with a layer of thickness $L_\mathrm{w}$, $d\ll L_\mathrm{w}\ll L$ containing it.
This layer is equivalent to the flat wall: they have the same jump of polarization and the same free charge, which create the same electric field in the bulk of the domains.
The virtual displacement of the walls will lead to the same changes of the charges on electrodes and the same changes in the fraction of the domains.
Thus the total forces acting on the zig-zag wall and the flat wall should be the same in the limit of $d/L\ll 1.$
This enables us to use Eq. \eqref{force_simplified} for the calculation the force acting in the walls of this type.

%
%
%%%%%%%%%%%%%%%%%%%%%%%%%%%%%%%%%               2222222
%                                                                                                                                       2      2
%                                                                                                                                       2      2
%                                                                                                                                       2      2
%%%%%%%%%%%%%%%%%%%%%%%%%%%%%%%%%               2222222

\section{Force on a domain wall between finite domains}\label{Force}

\begin{figure}[h]
\begin{center}
\includegraphics[width=0.5\textwidth]{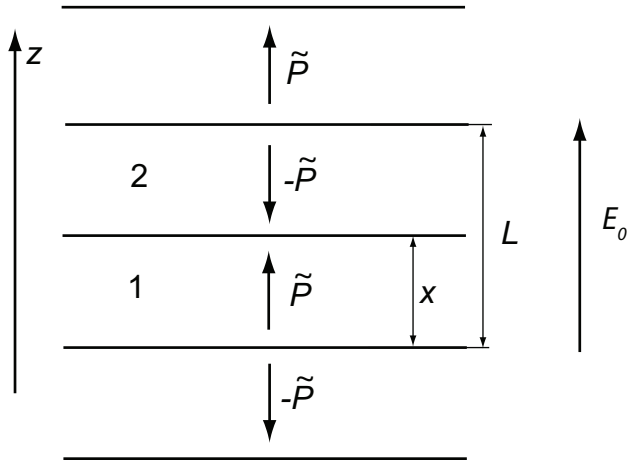}
\caption{Schematic of the periodic structure of "head-to-head" and "tail-to-tail" 180-degree domain walls with period $L$.
Domain walls are perpendicular to the polarization inside the domains $\tilde{P}$, which is different from the spontaneous polarization due to the depolarizing field.
 }
\label{periodic}
\end{center}
\end{figure}

In the previous section we obtained the force acting on the domain wall, carrying an arbitrary free charge density $\sigma_\mathrm{f}$.
The calculation of such force requires the knowledge of this charge.
An essential interesting feature of this problem is that it is dependent, in a unique way, on other parameters of the system.
This makes it possible to calculate the force, acting on a charged wall in a domain pattern, due to the application of an external electric field.
We will illustrate this for the case of a periodic domain pattern with  charged walls (see Fig. \ref{periodic}) in a uniaxial ferroelectric with the second-order phase transition.
We will be interested in the situation, typical for ferroelectrics\cite{Gureev}, where the screening of the bound charge in the wall is in the nonlinear regime, i.e. the carrier concentration inside the domain wall is much higher, than the homogeneous carrier concentration in the material.
%Such a structure was observed in the series of experiments by Shur et al. \cite{Shur93,Shur} in Pb$_5$Ge$_3$O$_{11}$.
We will consider a pattern with domains with the equal thicknesses $x=L/2$, where $L$ is the structure period,
having planar walls perpendicular to the spontaneous polarization.
%The same consideration is valid for the case of the zig-zag wall with thickness $d$ much smaller than the domain size (see Sect. \ref{sec:Generalized}).

As was shown in Refs. \cite{Guro68}, \cite{Gureev}, the charged wall in a finite sample is not completely compensated.
The net charge on the wall is proportional (due to the Gauss law) to the electric field inside the adjacent domains.
In the absence of external bias field, in a pattern containing charged domain walls, this field is not zero.
For the considered situation, where the screening regime in the wall is  nonlinear this field can be readily estimated.
Indeed,
in this regime, the free carrier density, needed for the polarization screening, can be provided exclusively by a band bending.
Obviously, such band bending implies a  voltage difference between the "head-to-head" and "tail-to -tail" domain walls,
which is  about $E_\mathrm{g}/q$, where $E_\mathrm{g}$ is the band-gap of the ferroelectric, $q$ is the absolute value of the electron charge \cite{Ivanchik,Gureev}.
Finally, the corresponding electric field inside a domain of the thickness $L/2$ is
\begin{equation}\label{field_domain}
\tilde{E}=\frac{2E_\mathrm{g}}{qL}.
\end{equation}
It can actually also be viewed as a partially compensated depolarizing field.
Due to this field the polarization inside the domains is reduced, compared to the nominal  spontaneous polarization of the ferroelectric.
For a ferroelectric with second-order phase transition,
using equation of state $E=\alpha P+ \beta P^3$,
the reduced polarization $\tilde{P}$ can be found as the negative solution to the following equation:

\begin{equation}\label{polarization_domain}
\alpha \tilde{P}+ \beta \tilde{P}^3=\frac{2E_\mathrm{g}}{qL}.
\end{equation}

%%%%%%%%%%%%%%%%%%%%%%%%%%%%%%%%%%%
%2AAAAAAAAAAAAAAAAAAAAAAAAAAAAAAAAAAAAAAAAAAAA

\subsection{Pressure on the wall between domains of equal sizes}\label{equal_domains}

For the configuration considered, the force component parallel to the wall, given by Eq. \eqref{force_parallel_local}, is zero,
whereas the pressure, given by Eq. \eqref{pressure}, is non-zero.
Considering domain wall between domains 1 and 2, one can rewrite Eq. \eqref{pressure} as follows:
\begin{equation}\label{pressure1}
p=\Phi_2-\Phi_1-\frac{1}{2}(E_1+E_2)(P_2-P_1-\sigma_\mathrm{f}),
\end{equation}
where $\Phi=(\alpha/2)P^2+(\beta/4)P^4$.
Here, the electric field and polarization are considered  positive if corresponding vectors are directed along the z axis (Fig. \ref{periodic}),
i.e. from domain 1 to domain 2.
It is clear from the structure of Eq.\eqref{pressure1} that the absolute values of the pressure, acting on the head-to-head and  tail-to-tail walls, are equal, whereas the corresponding forces are acting in opposite directions.
Thus, it suffices to calculate the pressure on the head-to-head walls between domains.

The Poisson equation for such configuration has the form (see Eq.\eqref{D} for the definition of the electric displacement) :
\begin{equation}\label{free_charge}
\sigma_\mathrm{f}=\varepsilon_\mathrm{b} E_2 + P_2 -\varepsilon_\mathrm{b} E_1-P_1,
\end{equation}
where electric fields inside the domains can be calculated using the equations of state:
\begin{equation}\label{state1}
E_1=\alpha P_1+ \beta P_1^3,
\end{equation}
\begin{equation}\label{state2}
E_2=\alpha P_2+ \beta P_2^3.
\end{equation}
The external electric field $E_0$ defined as $E_0=U/h$ can be expressed in terms of $E_1$ and $E_2$:
\begin{equation}\label{E0}
E_0=(E_1+ E_2)/2.
\end{equation}

Using Eqs.~\eqref{free_charge}-\eqref{E0} we can present the polarizations inside the domains as a Taylor expansion with respect to $E_0$ as follows:
\begin{equation}\label{polarization_expansion1}
P_1=\tilde{P}+\chi E_0+\eta E_0^2,
\end{equation}
\begin{equation}\label{polarization_expansion2}
P_2=-\tilde{P}+\chi E_0-\eta E_0^2,
\end{equation}
where
\begin{equation}\label{ksi}
\chi=\frac{1}{\alpha+3\beta\tilde{P}^2}
\end{equation}
is the susceptibility of the ferroelectric at $P=\tilde{P}$ and
\begin{equation}\label{eta}
\eta=\frac{-3\beta\tilde{P}}{(\alpha+3\beta\tilde{P}^2)^2(\alpha+3\beta\tilde{P}^2+1/\varepsilon_\mathrm{b})}=\frac{-3\beta\tilde{P}\chi^2}{1/\chi+1/\varepsilon_\mathrm{b}}.
\end{equation}

Substituting polarization from Eqs. \eqref{polarization_expansion1}, \eqref{polarization_expansion2} into Eq.\eqref{pressure1} and keeping terms up to third power with respect to $E_0$, one finds:
\begin{equation}\label{pressure_central}
p=\frac{\varepsilon_\mathrm{f}}{\varepsilon_\mathrm{b}}(2\tilde{P}+\sigma_\mathrm{f})E_0-2\beta\tilde{P}\chi^3E_0^3,
\end{equation}
where
\begin{equation}\label{epsilonf}
\varepsilon_\mathrm{f}=\varepsilon_\mathrm{b}+\chi
\end{equation}
is the permittivity of the ferroelectric at $P=\tilde{P}$.

From the Poisson equation, Eq. \eqref{free_charge}, we can find:
\begin{equation}\label{E_domain}
2\tilde{P}+\sigma_\mathrm{f}=2\varepsilon_\mathrm{b}\tilde{E},
\end{equation}
where $\tilde{E}$ is the electric field inside the domain, which in turn can be found from  Eq. \eqref{field_domain}.
Thus we obtain:
\begin{equation}\label{nccharge}
2\tilde{P}+\sigma_\mathrm{f}=\frac{4\varepsilon_\mathrm{b}E_\mathrm{g}}{qL}.
\end{equation}
Using Eq.\eqref{nccharge}, we rewrite Eq.~\eqref{pressure_central} as follows:
\begin{equation}\label{pressure_thickness}
p=\frac{4\varepsilon_\mathrm{f}E_\mathrm{g}}{qL}E_0-2\beta\tilde{P}\chi^3E_0^3.
\end{equation}
Note that in Eq.~\eqref{pressure_thickness} the $\chi$ and $\varepsilon_\mathrm{f}$ should be calculated for the reduced values of the polarization inside domains.

It is instructive to compare the part of the pressure, linear in $E_0$,  given by Eq.\eqref{pressure_thickness}  for the case where the  polarization inside the domains is close to the bulk spontaneous polarization $P_0$ with the pressure on the neutral domain wall
$2P_0E_0$.
Physically, the condition of the polarization close to $P_0$ means that the depolarizing field inside domains, $\tilde{E}$, is much smaller than the thermodynamic coercive field $E_\mathrm{coer}=(2/(3\sqrt{3}))|\alpha|P_0$.
In this case, neglecting the difference between $\tilde{P}$  and $P_0$ and using Eqs. \eqref{pressure_thickness}, \eqref{ksi} one finds $p/(2P_0E_0)\approx 0.2 \tilde{E}/E_\mathrm{coer}$.
Since, in the situation considered, the depolarizing field inside the domain $\tilde{E}$ is much smaller than $E_\mathrm{coer}$, the pressure on the charged wall is expected to be much smaller than the pressure on the neutral wall.

The situation, where the depolarizing field  approaches the thermodynamic coercive field, corresponds to a polarization instability inside domains manifesting itself in a divergency of the permittivity $\varepsilon_\mathrm{f}$.
This formally implies (via Eqs. \eqref{pressure_thickness} and \eqref{field_domain}) an unlimited  increase of the pressure acting on the wall when the domain period $L$ approaches a certain critical value.
This academically interesting situation is, however, hardly reachable experimentally \cite{minimal_period}. 
For this reason, in this paper we will further discuss only the situation where the ferroelectric inside domains is far from the aforementioned instability.
On the practical level, this means that in further estimates we can neglect the difference between the value of the polarization inside domain, $\tilde{P}$, and the spontaneous polarization of the ferroelectric $P_0$.

%%%%%%%%%%%%%%%%%%%%%%%%%%%%%%%%%%%%%%%%%%POLING%%%%%%%%%%%%%%%%%%%%%%%%%%%%%%%%%%
\subsection{Polling a sample with charged domain walls}\label{linear}

In this section we will consider the problem of poling a sample, where the periodical structure with charged domain walls shown in Fig. \ref{periodic} was formed.
We will consider the case of a small external field, where we can consider only the linear in $E_0$ part of the pressure.
Thus, keeping only the linear term in Eq. \eqref{pressure_thickness}, we find this pressure in the form:
\begin{equation}\label{pressure_linear}
p=\frac{4\varepsilon_\mathrm{f}E_\mathrm{g}}{qL}E_0.
\end{equation}

For the model developed in the previous section and used in the present one, the free carrier concentration in the domain wall is close to metallic.
This means that, for a realistic concentration of the trapping centers, the main part of the free carriers in the wall is not trapped, having  high mobility and, thus, being able to easily follow the domain wall in its motion.
A realistic situation, which we consider, is that where the time needed for the free-charge to transfer between the domain walls (which depends on the conductivity of the material and domain size) is large in comparison to the time of poling.
That means that in our model, when a domain wall moves away from its position in a periodic pattern,  the free charge which its carries does not change.

Defects, which always exist in real materials, lead to the domain wall pinning.
The pinning acts as dry friction in mechanics.
When we apply a small pressure on the domain wall, its position is stabilized because of the pinning.
To start the domain wall motion, the applied pressure, given by Eq.\eqref{pressure_linear},  should be equal to the maximal "stabilizing" pressure of the pinning $p_\mathrm{pinning}$:
\begin{equation}\label{pressure_pinning}
p=p_\mathrm{pinning}.
\end{equation}

Equations \eqref{pressure_pinning} and \eqref{pressure_linear} enable us to find the electric field needed to start the motion of the walls from the initial position in the equidistant domain pattern.
To start the domain wall motion, it is necessary, but not sufficient for poling.
To be sure that condition \eqref{pressure_pinning} corresponds to the onset of successful poling, we should check if the pressure $p$ is greater or equal to the pinning pressure at any position of the wall after it starts moving.
If the pining pressure is independent of the position of the domain wall, successful poling will occur if, at given external electric field, the pressure acting at the wall in a shifted position  is  larger than that  in the original position.
In reality, one expects the pining pressure to be larger in the original position because of the defect accumulations at the wall during the time before the poling; the situation observed experimentally \cite{Shur}.
It is clear that for this situation the aforementioned criterium of successful poling holds as well.

Let us check if the criterium of successful poling is met for our system, shown in Fig. \ref{periodic}.
Since  the absolute value of the pressure acting on the head-to-head and  tail-to-tail walls are equal, whereas the corresponding forces are acting in the opposite directions, the period of the pattern will not be affected by the application of an electric field. 
Thus, when discussing poling we can address just one head-to-head wall moving between two tail-to-tail walls spaced by a time-independent distance $L$ (see Fig. \ref{periodic}).   

Let us first consider how the polarizations $P_1$, $P_2$ and electric fields $E_1$, $E_2$ inside the domains will change after a domain wall displacement from initial position.
Differentiating the Poisson equation, Eq. \eqref{free_charge}, and taking into account that the free charge on the wall remains constant we get:
\begin{equation}\label{poisson_dif}
\left( \varepsilon_\mathrm{b}\frac{\partial E_2}{\partial P_2}+1 \right) \frac{\partial P_2}{\partial x} = \left( \varepsilon_\mathrm{b}\frac{\partial E_1}{\partial P_1}+1 \right) \frac{\partial P_1}{\partial x}.
\end{equation}
Here $\partial E/\partial P\approx \chi^{-1} \approx \varepsilon_\mathrm{f}^{-1}$, thus $\varepsilon_\mathrm{b} (\partial E/\partial P) \approx \varepsilon_\mathrm{b}/\varepsilon_\mathrm{f}\ll1$ and we can conclude that 
\begin{equation}\label{dP12}
\frac{\partial P_1}{\partial x} \approx \frac{\partial P_2}{\partial x}.
\end{equation}
The sign of the derivatives in Eq. \eqref{dP12} can be obtained from the relation between  
 average field $E_0$ applied to the ferroelectric and fields inside the domains $E_1$ and $E_2$: 
\begin{equation}\label{Field_E1_E2}
E_1x+E_2(L-x)=E_0L.
\end{equation}
Differentiating this equation we obtain:
\begin{equation}\label{d_E1_E2}
\frac{\partial E_1}{\partial x} x+\frac{\partial E_2}{\partial x} (L-x)=-E_1+E_2.
\end{equation}
In the case of head-to-head wall considered here (see Fig. \ref{periodic}) $E_1<0$, $E_2>0$, implying that the expression in the right hand side in Eq. \eqref{d_E1_E2} is positive.
Thus, the left hand side expression in Eq.  \eqref{d_E1_E2} is also positive and we obtain
\begin{equation}\label{d_E1_E2_more0}
\frac{\partial E_1}{\partial x} x+\frac{\partial E_2}{\partial x} (L-x)=\frac{\partial E_1}{\partial P_1}\, \frac{\partial P_1}{\partial x}\, x+\frac{\partial E_2}{\partial P_2}\, \frac{\partial P_2}{\partial x}\,  (L-x)\, > 0.
\end{equation}
Here $\partial E_1/\partial P_1$ and $\partial E_2/\partial P_2$ are inverse susceptibilities (which are positive);
domain widths $x$, $L-x$ are also positive. 
Thus, from Eq. \eqref{d_E1_E2_more0} using Eq. \eqref{dP12}  it follows 
\begin{equation}\label{pressure_sign}
\begin{split}
\partial P_1/\partial x\approx \partial P_2/\partial x>0.
\end{split}
\end{equation}
Below we will consider the positive displacement of the domain wall, i.e. the size of the domain 1 $x$ is increasing during this displacement.
This situation corresponds to the external field applied from domain 1 to domain 2 ($E_0>0$). 
For this case, using Eq.~\eqref{pressure_sign} we can conclude that polarization in domain 1, $P_1>0$, and it will increase (the absolute value of depolarizing field $E_1$ in this domain will decrease) while in the domain 2, $P_2<0$, and its absolute value will decrease (the absolute value of the field $E_2$ will increase).
We can summarize the result for polarization change in a compact form as:
\begin{equation}\label{P_moving}
|P_2|<\tilde{P}<|P_1|<P_0,
\end{equation}
\begin{equation}\label{E_moving}
|E_1|<|\tilde{E}|<|E_2|.
\end{equation}

Now, using Eqs. \eqref{pressure1}, \eqref{free_charge}, we can obtain the pressure 
 on the domain wall, when it is in an arbitrary position, as:
\begin{equation}\label{pressure_periodic}
p=\Phi_2-\Phi_1+\frac{\varepsilon_\mathrm{b}}{2}(E_2^2-E_1^2). 
\end{equation}
In view of Eq. \eqref{E_moving} when the domain wall moves in the positive sense of the $z$-axis ($x>1/2$), $E_2^2$ increases, $E_1^2$ decreases, thus the last term in Eq. \eqref{pressure_periodic} increases.
Thus, to show that the pressure increases when the wall is displaced from the central position it is enough to check that term $\Phi_2-\Phi_1$ increases.
Using Taylor expansion of polarization similar to Eqs. \eqref{polarization_expansion1}-\eqref{ksi} and keeping only linear terms with respect to $E_0$  we find:
\begin{equation}\label{pressure_x}
\Phi_2-\Phi_1=G(\tilde{P}_1,\tilde{P}_2)+H(\tilde{P}_1,\tilde{P}_2)E_0,
\end{equation}
where
\begin{equation}\label{G}
G(\tilde{P}_1,\tilde{P}_2)=\frac{1}{4}\beta(\tilde{P}_1^2-\tilde{P}_2^2)(2P_0^2-\tilde{P}_2^2-\tilde{P}_1^2),
\end{equation}
\begin{equation}\label{H}
H(\tilde{P}_1,\tilde{P}_2)=\frac{\tilde{P}_2(\tilde{P}_2^2-P_0^2)}{3\tilde{P}_2^2-P_0^2}-\frac{\tilde{P}_1(\tilde{P}_1^2-P_0^2)}{3\tilde{P}_1^2-P_0^2}.
\end{equation}
Here $\tilde{P}_1$, $\tilde{P}_2$, $\tilde{E}_1$, $\tilde{E}_2$ are polarizations and electric fields in non-symmetric situation ($x\neq L-x$) at zero external field in the domains 1 and 2, respectively.

In the case of the zero external field Eq.~\eqref{P_moving} leads to $\tilde{P}_2<\tilde{P}_1<P_0$,
and it immediately follows that $G(\tilde{P}_1,\tilde{P}_2)>0$.
Thus if we displace the domain wall from the central position, and the external field is zero, the force will act on the wall, which is parallel to the displacement.
Thus the equidistant periodic structure with charged domain walls is unstable.
It corresponds to a maximum of the energy.
In real systems, the position of charged domain walls are stabilized because of the pinning on defects, discussed above.

Now we will analyze the second term in Eq. \eqref{pressure_x}, which is proportional to $E_0$.
Using Eqs. \eqref{H}, \eqref{dP12} we find 
\begin{equation}\label{dH}
\frac{\partial H}{\partial x}\approx 
6(\tilde{P}_1^2-\tilde{P}_2^2)\frac{P_0^4(\tilde{P}_1^2+\tilde{P}_2^2-P_0^2)+3P_0^2\tilde{P}_1^2\tilde{P}_2^2}{(3P_2^2-P_0^2)^2(3P_1^2-P_0^2)^2}\, \frac{\partial P}{\partial x}.
\end{equation}

From Eq. \eqref{dH} using Eqs. \eqref{pressure_sign}, \eqref{P_moving} and taking into account that $\tilde{P}_1\approx P_0$, $\tilde{P}_2\approx P_0$ we have $\partial H/\partial x>0$.

Thus after we displaced domain wall from the central position by application of the external electric field,
the pressure acting on the wall will increase, and the criterium of successful poling is met. 
This means that the switching is governed by the condition in the original position of the domain wall given by Eqs.  \eqref{pressure_pinning} and \eqref{pressure_linear}.

%%%%%%%%%%%%%%%%%%%%%%%%%%%%%%%%%%%%%%%%%%%%%NON-LINEAR%%%%%%%%%%%%%%%%%%%%%%%%%%%%%%%%%%
%%%%%%%%%%%%%%%%%%%%%%%%%%%%%%%%%%%%%%%%%%%%%%%%%%%%%%%%%%%%%%%%%%%%%%%%%%%%%%%%%%%%%%
\subsection{Nonlinear part of the pressure}\label{nonlinear}

The nonlinear part of the pressure Eq. \eqref{pressure_central} is directed in the opposite direction to the linear part, and, for the structure with large domains where $\tilde{P}\to P_0$, it leads to the nonlinear pressure obtained as \cite{Mokry}:
\begin{equation}\label{pressure_fully}
p=-(\chi^2/P_0)E_0^3.
\end{equation}
If the domain wall is well compensated and the applied field is high enough, the nonlinear pressure can play the decisive role and resultant pressure can act opposite to the linear one.
Thus, if we abruptly apply high voltage, the domain wall may move in the direction, opposite to the direction of motion at low voltage.
Let's estimate the electric field at which the values of the nonlinear and linear pressure are equal.
Setting $\tilde{P}\approx P_0$ from Eq. \eqref{pressure_thickness} one can find:
\begin{equation}\label{E_NL}
E_0=\sqrt{4E_\mathrm{g}P_0/qL\chi}.
\end{equation}
 For the typical values $E_\mathrm{g}=3$ eV, $P_0=0.1$ C/m$^2$, $\chi=10^3\times\varepsilon_0$, $L=10$ $\mu$m (this is about the period of the structure observed by Shur \cite{Shur}) one can check that the corresponding field inside the domain is far from the thermodynamic coercive field. 
 Thus Eq. \eqref{E_NL} can be used and it leads to $E_0\approx 40$ kV/cm.
 This field is decreasing with increasing the domain size.

 %%%%%%%%%%%%%%%%%%%%%%%%%%%%%
  \section{Discussion and Conclusions}\label{Discussion}

  We obtained a general formula for the local force acting on the domain wall with arbitrary free charge on it.
  In general, this force in not perpendicular to the domain wall.
   The force acting on the domain wall depends on the compensating free charge on it.
  The better the compensation, the smaller the force acting on the domain wall.
  The force acting on the fully compensated wall is zero.

  The compensating free charge on the domain wall depends on the characteristic size of the domain structure.
  We considered the periodic domain structure with parallel "head-to-head" and "tail-to-tail" walls,
  and obtained the force acting on the domain wall as a function of the structure period.
Smaller domain size leads to a larger force.
  For the structure which is not very dense the force acting on the charged domain wall is much less than the force acting on the neutral domain wall.
  Thus poling of the sample with such a structure is difficult in comparison with poling of the sample with neutral domain walls.
  This result is in good agreement with the experimental data obtained by Shur et al.\cite{Shur}

  We also showed that in an ideal crystal, the periodic structure with charged domain walls is unstable.
  It corresponds to the maximum of the energy.
  Nevertheless, in real materials it can be stabilized due to pinning provided by defects.

  \section{Acknowledgments}\label{Acknowledgments}

One of the authors (P.M.) acknowledges the support from the Czech Science Foundation project No. GACR P204/10/0616.

 %%%%%%%%%%%%%%%%%%%%%%%%%%%%%%

 %%%%%%%%%%%%%%%%%%%%%%%%%%%%%%%%%
 %
\appendix
  \section{Derivation of the local force density acting on the element of the domain wall}\label{appendix_force}

\begin{figure}
    \includegraphics[width=85mm]{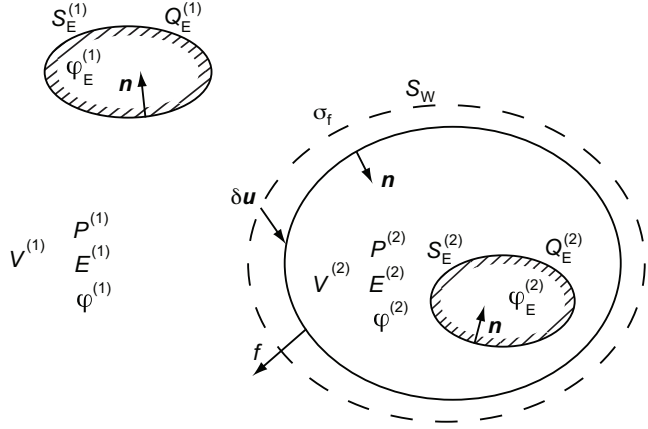}
    \caption{General system of a ferroelectric in contact with two conductors.
Domain wall $S_W$ separates the ferroelectric into two domains of
volumes $\1{V}$ and $\2{V}$. There are two conductors, which carry
charges $\1{Q_E}$ and $\2{Q_E}$ and have electric potentials
$\1{\varphi_E}$ and $\2{\varphi_E}$ within each domain. The
quantities within each domain are the ferroelectric part of polarization $\1{P_i}$ and
$\2{P_i}$, electric field $\1{E_i}$ and $\2{E_i}$ and electric
potential $\1{\varphi}$ and $\2{\varphi}$. Symbol $\delta u$
stands for the virtual displacement of the domain wall and symbol
$f$ stands for the local force density of external sources on the
domain wall in order to keep the system in equilibrium.}
    \label{local}
\end{figure}

 We analyze the local pressure on the ferroelectric
domain wall in a system shown in Fig. \ref{local}.
We consider that inside a ferroelectric material with
ferroelectric part of polarization $\1{P_i}$
there exists a closed domain of the material in a different
polarization state $\2{P_i}$.
The closed boundary splits the
ferroelectric into two domains, which have volumes $\1{V}$ and
$\2{V}$, respectively.
Inside domains there are
conductors, which carry charges $\1{Q_E}$ and $\2{Q_E}$,
respectively, which are at electric potentials $\1{\varphi_E}$ and
$\2{\varphi_E}$, respectively.
We consider that the bound charges due
to discontinuous change of polarization at the domain wall $S_W$
are partially compensated by free charges of surface density
$\sigma_f$. The charges on conductors and on the domain wall
produce electric fields $\1{E_i}$ and $\2{E_i}$ within each
domain. Symbols $\1{\varphi}$ and $\2{\varphi}$ stand for the electric
potentials within each domain.

In order to obtain a general formula for the local pressure
on domain wall, we follow the principle of virtual displacements, formulated in Sect.\ref{sec:Generalized}.
Instead of Eq. \eqref{G_full_comp} we will use
the thermodynamic function $G$ in a more general form:
\begin{equation}\label{G}
    G = \int_V \left[\Phi(P) + \frac 12\varepsilon_\mathrm{b} E^2\right]\, dV - \sum_i\varphi_E^{(i)}\, Q_E^{(i)},
\end{equation}
where the first term represents the energy associated with the
ferroelectric part of polarization, $\Phi(P)$, and
the electric field energy integrated over the volume $V$ of
ferroelectric, and the second term represents the subtracted work
of the electric sources.

In what follows it is convenient to transform the work of electric
sources into volume integrals:
\begin{subequations}
\label{eq:07:IntIds}
\begin{eqnarray}
    \label{eq:07a:IntId1}
    \varphi_E^{(1)}\, Q_E^{(1)} &=& - \varphi_E^{(1)}\,
    \int_{S_E^{(1)}}D_i^{(1)} n_i\, dS =
    \int_{V^{(1)}} E_i^{(1)} D_i^{(1)}\, dV
        +
        \int_{S_W} \varphi^{(1)} D_i^{(1)} n_i\, dS, \\
    \label{eq:07b:IntId1}
    \varphi_E^{(2)}\, Q_E^{(2)} &=& - \varphi_E^{(2)}\,
    \int_{S_E^{(2)}}D_i^{(2)} n_i\, dS = \int_{V^{(2)}} E_i^{(2)} D_i^{(2)}\, dV
        -
        \int_{S_W} \varphi^{(2)} D_i^{(2)} n_i\, dS,
\end{eqnarray}
\end{subequations}
where $\1{D_i}$ and $\2{D_i}$ are the vectors of electric
displacement in domains $\1{V}$ and $\2{V}$, respectively, and
where we considered the absence of free charges, i.e. $\mathrm{div} D_i=0$,
inside domains $V^{(1)}$ and $V^{(2)}$ and vanishing the surface
integral over the domain $V^{(1)}$ at infinity. Orientations of the
normal vectors $n_i$ are indicated in Fig. \ref{local}. Since the
bulk quantities, e.g. $\varphi D_i$, are generally discontinuous
at the domain wall $S_W$, it will be useful to denote their jump
by, e.g.
\begin{equation}
    \DifrOp{\varphi D_i} = \2{\varphi}\2{D_i} - \1{\varphi}\1{D_i}.
    \label{eq:08:DefOpDifrence}
\end{equation}
If we employ the expression for electric displacement
$D_i=P_i+\varepsilon_\mathrm{b}\,E_i$, the thermodynamic function $G$ can
be expressed in a form:
\begin{eqnarray}
    \label{eq:09:G}
    G &=& \int_{\1{V}}\left[
        \1{\Phi}(\1{P}) - \frac 12 \varepsilon_\mathrm{b}\1{E_i}\1{E_i}
        - \1{E_i}\1{P_i}
    \right]\, dV + \\
    \nonumber
    &+& \int_{\2{V}}\left[
        \2{\Phi}(\2{P}) - \frac 12 \varepsilon_\mathrm{b}\2{E_i}\2{E_i}
        - \2{E_i}\2{P_i}
    \right]\, dV + \\
    &+& \int_{S_W}\DifrOp{
        \varphi\left(P_i + \varepsilon_\mathrm{b}\,E_i\right)
    }\, n_i\, dS.
    \nonumber
\end{eqnarray}

Now our task is to express the variation of thermodynamic function
$\delta G$, which is produced by the virtual displacement of
domain wall. When doing this, we have to keep in mind that the
bulk quantities $\varphi$, $E_i$, and $P_i$ are not independent
variables. The relation between the electric field and the
electric potential is given by $E_i=-\partial\varphi/\partial x_i$
and one has to consider the continuity of electric displacement at
the domain wall, i.e. $\DifrOp{P_i + \varepsilon_\mathrm{b}\,E_i}\, n_i =
\sigma_f$. These additional conditions can be introduced into our
treatment by adding new variables $\lambda_i$ and $\mu$ according
to the method of Lagrange multipliers. Using this approach we
obtain a new thermodynamic function $L=L(\varphi, E_i, P_i,
\lambda_i, \mu)$, where \emph{all bulk variables in the both
domains are considered independent}:
\begin{eqnarray}
    \label{eq:10:L}
    L &=& \int_{\1{V}}\left[
        \1{\Phi}(\1{P}) - \frac 12 \varepsilon_\mathrm{b}\1{E_i}\1{E_i}
        - \1{E_i}\1{P_i} + \1{\lambda_i}\, \left(\1{E_i} + \frac{\partial\1{\varphi}}{\partial x_i}\right)
    \right]\, dV + \\
    \nonumber
    &+& \int_{\2{V}}\left[
        \2{\Phi}(\2{P}) - \frac 12 \varepsilon_\mathrm{b}\2{E_i}\2{E_i}
        - \2{E_i}\2{P_i} + \2{\lambda_i}\, \left(\2{E_i} + \frac{\partial\2{\varphi}}{\partial x_i}\right)
    \right]\, dV + \\
    &+& \int_{S_W}\Bigl[
        \DifrOp{\varphi\left(P_i + \varepsilon_\mathrm{b}\,E_i\right)}\, n_i
        + \mu\left(\DifrOp{P_i + \varepsilon_\mathrm{b}\,E_i}\, n_i - \sigma_f\right)
    \Bigr]\, dS.
    \nonumber
\end{eqnarray}
The virtual displacement $\delta u$ of domain wall, which in general is not normal
to the wall surface, shown in Fig. \ref{local} produces the variation of all
independent quantities $\delta\varphi$, $\delta E_i$, $\delta
P_i$, $\delta\lambda_i$, and $\delta\mu$; except the external
force $f$ on the domain wall and the electric potential on
conductors $\1{\varphi_E}$ and $\2{\varphi_E}$, which are constant
in the system. The variation of quantities results in the
variation of thermodynamic function $\delta L$. In equilibrium the
variation $\delta L$ equals the work produced by external force
 during the virtual displacement of domain wall
\begin{equation}
    \label{eq:11:deltaWp}
    \delta W= - \int_{S_W}f_i\delta u_i\, dS,
\end{equation}
where $f_i$ is the local force density per unit area of the domain wall, $S_W$ is the area of domain wall.

In order to express the variation $\delta L$, it is useful to
consider following formula for the variation of volume integral
\begin{subequations}
\label{eq:12:Var}
\begin{equation}
    \delta\left\{\int_{V}g\, dV\right\} =
        \int_{V} \delta g\, dV
        \pm
        \int_{S_W} g\,\delta u_k n_k\, dS,
    \label{eq:12a:VarV}
\end{equation}
where $g$ is an arbitrary bulk quantity and plus and minus correspond
to volume $\1{V}$ and $\2{V}$, respectively. The physical
interpretation of the above formula is that the variation of
volume integral includes the contribution due to variation of bulk
quantity $\delta g$ and the contribution due to the volume change
of domains produced by the virtual displacement of domain wall
$\delta u$. Similar formula can be written for the variation of
surface integral over the domain wall
\begin{equation}
    \delta\left\{\int_{S_W}g\, dS\right\} =
        \int_{S_W} \delta g\, dS
        +
        \int_{S_W}
        \frac{\partial g}{\partial x_k} \, \delta u_k\, dS,
    \label{eq:12b:VarSW}
\end{equation}
\end{subequations}
where the first term on the right-hand side represents the
contribution due to variation of bulk quantity $\delta g$ on the
domain wall and the second term represents the contribution due to
change of the domain wall position during the virtual
displacement.

Applying Eqs. (\ref{eq:12:Var}) to function $L$ given by Eq.
(\ref{eq:10:L}), we obtain
\begin{eqnarray}
    \label{eq:13:deltaL}
    \delta L &=& \int_{\1{V}}\left\{
        \delta \1{P_i} \left[
            \frac{\partial\1{\Phi}}{\partial P_i} - \1{E_i}
        \right]
        -
        \delta\1{E_i} \left[
             \1{P_i} + \varepsilon_\mathrm{b}\1{E_i} - \1{\lambda_i}
        \right]
        + \right. \\
    \nonumber
    \lefteqn{
        \left. \hspace{2.5cm}
        +
        \frac{\partial \delta \1{\varphi}}{\partial x_i} \1{\lambda_i}
        +
        \delta \1{\lambda_i} \left[
            \1{E_i} + \frac{\partial\1{\varphi}}{\partial x_i}
        \right]
    \right\} dV + }\\
    \nonumber
    &+& \int_{\2{V}}\left\{
        \delta \2{P_i} \left[
            \frac{\partial\2{\Phi}}{\partial P_i} - \2{E_i}
        \right]
        -
        \delta\2{E_i} \left[
            \2{P_i} + \varepsilon_\mathrm{b}\2{E_i} - \2{\lambda_i}
        \right]
        + \right. \\
    \nonumber
    \lefteqn{
        \left. \hspace{2.5cm} +
        \frac{\partial \delta \2{\varphi}}{\partial x_i} \2{\lambda_i}
        +
        \delta \2{\lambda_i} \left[
            \2{E_i} + \frac{\partial\2{\varphi}}{\partial x_i}
        \right]
    \right\} dV + } \\
    \nonumber
    &+& \int_{S_W}\left\{
        \delta\2{\varphi}\left[\2{P_i} + \varepsilon_\mathrm{b}\2{E_i}\right]n_i
        - \delta\1{\varphi}\left[\1{P_i} + \varepsilon_\mathrm{b}\1{E_i}\right] n_i
        + \right. \\
    \nonumber
    \lefteqn{  \hspace{2.5cm}
        +\left(\delta\2{P_i} + \varepsilon_\mathrm{b}\delta\2{E_i}\right)n_i
        \left(\2{\varphi}+\mu\right)
        -\left(\delta\1{P_i} + \varepsilon_\mathrm{b}\delta\1{E_i}\right)n_i
        \left(\1{\varphi}+\mu\right) -
    } \\
    \nonumber
    \lefteqn{ \hspace{2.5cm}
        \left.
        + \delta\mu\left(\DifrOp{P_i + \varepsilon_\mathrm{b}E_i} n_i -
        \sigma_f\right) +
    \right. }\\
    \nonumber
    \lefteqn{  \hspace{2.5cm}
        - \delta u_k \left(
            \DifrOp{\Phi(P) - \frac 12 \varepsilon_\mathrm{b}E_iE_i
                - E_iP_i + \lambda_i\, \left(E_i + \frac{\partial\varphi}{\partial x_i}\right)
            }n_k +
        \right.
    } \\
    \lefteqn{  \hspace{2.5cm}
            \left. \left. +
            \frac\partial{\partial x_k}
            \left(
            \DifrOp{
                \varphi\left(P_i + \varepsilon_\mathrm{b}E_i\right)n_i
            } +
                \mu\left(\DifrOp{P_i + \varepsilon_\mathrm{b}\,E_i}\, n_i - \sigma_f\right)
            \right)
        \right)
    \right\}\, dS.
    }
    \nonumber
\end{eqnarray}
The above formula can be further transformed using following
integral expressions:
\begin{subequations}
\label{eq:14:IntegEq}
\begin{eqnarray}
    \int_{\1{V}}\1{\lambda_i}\left(\frac{\partial\delta\1\varphi}{\partial x_i}\right) dV
    +
    \int_{\1{V}}\left(\frac{\partial\1{\lambda_i}}{\partial x_i}\right)\delta\1\varphi
    dV &=&
    \int_{S_W}\1{\lambda_i}\delta\1{\varphi}n_i dS,
%    + \int_{\1{S_E}}\1{\lambda_i}\delta\1{\varphi}n_i dS
    \label{eq:14a:IntegEq1} \\
    \int_{\2{V}}\2{\lambda_i}\left(\frac{\partial\delta\2\varphi}{\partial x_i}\right) dV
    +
    \int_{\2{V}}\left(\frac{\partial\2{\lambda_i}}{\partial x_i}\right)\delta\2\varphi
    dV &=&
    - \int_{S_W}\2{\lambda_i}\delta\2{\varphi}n_i dS,
%    + \int_{\2{S_E}}\2{\lambda_i}\delta\2{\varphi}n_i dS
    \label{eq:14b:IntegEq2}
\end{eqnarray}
\end{subequations}
where we consider that, during the virtual displacement of domain
wall, the electric potential on conductors is kept constant by the
electric sources, i.e. the variations of electric potentials on
conductors are zero, and that the surface integral over domain
$V^{(1)}$ is vanishing at infinity.

Applying Eqs. (\ref{eq:14:IntegEq}) to formula
(\ref{eq:13:deltaL}), we readily obtain the variation $\delta L$
in a form
\begin{eqnarray}
    \label{eq:16:deltaL}
    \delta L &=& \int_{\1{V}}\left\{
        \delta \1{P_i} \left[
            \frac{\partial\1{\Phi}}{\partial P_i} - \1{E_i}
        \right]
        -
        \delta\1{E_i} \left[
            \1{P_i} + \varepsilon_\mathrm{b}\1{E_i} - \1{\lambda_i}
        \right]
        -
    \right. \\
    \nonumber
    \lefteqn{\hspace{2.5cm}
        \left.
            -
            \delta \1{\varphi} \frac{\partial \1{\lambda_i}}{\partial x_i}
            +
            \delta \1{\lambda_i} \left[
                \1{E_i} + \frac{\partial\1{\varphi}}{\partial x_i}
            \right]
        \right\} dV +
    }\\
    \nonumber
    &+& \int_{\2{V}}\left\{
        \delta \2{P_i} \left[
            \frac{\partial\2{\Phi}}{\partial P_i} - \2{E_i}
        \right]
        -
        \delta\2{E_i} \left[
            \2{P_i} + \varepsilon_\mathrm{b}\2{E_i} - \2{\lambda_i}
        \right]
        -
    \right. \\
    \nonumber
    \lefteqn{\hspace{2.5cm}
    \left.
        -
        \delta \2{\varphi} \frac{\partial \2{\lambda_i}}{\partial x_i}
        +
        \delta \2{\lambda_i} \left[
            \2{E_i} + \frac{\partial\2{\varphi}}{\partial x_i}
        \right]
    \right\} dV +
    }\\
    \nonumber
    &+& \int_{S_W}\left\{
        \delta\2{\varphi}\left[\2{P_i} + \varepsilon_\mathrm{b}\2{E_i} - \2{\lambda_i}\right]n_i
        - \delta\1{\varphi}\left[\1{P_i} + \varepsilon_\mathrm{b}\1{E_i}- \1{\lambda_i}\right]
        n_i
        +
    \right.
    \\
    \nonumber
    \lefteqn{ \hspace{2.5cm}
        +\left(\varepsilon_\mathrm{b}\delta\2{E_i} + \delta\2{P_i}\right) n_i
        \left(\2{\varphi}+\mu\right)
        -\left(\varepsilon_\mathrm{b}\delta\1{E_i} + \delta\1{P_i}\right) n_i
        \left(\1{\varphi}+\mu\right) -
    }\\
    \nonumber
    \lefteqn{\hspace{2.5cm}
        \left.
        + \delta\mu\left(\DifrOp{P_i + \varepsilon_\mathrm{b}E_i} n_i -
        \sigma_f\right) +
    \right.
    }\\
    \nonumber
    \lefteqn{ \hspace{2.5cm}
        - \delta u_k \left(
            \DifrOp{
                \Phi(P)
                - \frac 12 \varepsilon_\mathrm{b}E_iE_i
                - E_iP_i
                + \lambda_i\, \left(E_i + \frac{\partial\varphi}{\partial x_i}\right)
            } n_k
            +
        \right.
    }\\
    \lefteqn{ \hspace{2.5cm}
        \left. \left.
            + \frac\partial{\partial x_k}
            \left(
                \DifrOp{
                    \varphi\left(P_i + \varepsilon_\mathrm{b}E_i\right)n_i
                }
                +
                \mu\left(
                    \DifrOp{P_i + \varepsilon_\mathrm{b}\,E_i}\, n_i
                    -
                    \sigma_f
                \right)
            \right)
        \right)
    \right\}\, dS,
    }
    \nonumber
\end{eqnarray}
which can be combined with the work of external pressure $\delta
W$ during virtual displacement of domain wall given by Eq.
(\ref{eq:11:deltaWp}). Employing the principle of virtual
displacements $\delta L=\delta W\,$ yields the formula for the
local force density of external sources on the domain wall
\begin{eqnarray}
    \label{eq:16:pIs}
   f_k &=& \DifrOp{
                \Phi(P) -
                \frac 12 \varepsilon_\mathrm{b} E_i E_i -
                E_i P_i +
                \lambda_i\, \left(
                    E_i + \frac{\partial\varphi}{\partial x_i}
                \right)
            } n_k -
    \\
    \nonumber
    \lefteqn{\hspace{2.5cm}
        - \frac {\partial}{\partial x_k}
        \left\{
               \DifrOp{
                    \varphi\left(P_i + \varepsilon_\mathrm{b}E_i\right)n_i
                }
                +
                \mu\left(
                    \DifrOp{P_i + \varepsilon_\mathrm{b}\,E_i}\, n_i
                    -
                    \sigma_f
                \right)
        \right\} ,
    }
\end{eqnarray}
equations of motion
\begin{equation}
    \label{eq:17:EqOfMotion}
    \frac{\partial\Phi}{\partial P_i} = E_i,
\end{equation}
electric displacement $\lambda_i = \varepsilon_\mathrm{b}E_i+P_i$, Gauss'
law for electric displacement $\partial\lambda_i/\partial x_i =
0$, relationship between electric field and electric potential
$E_i = -\partial\varphi/\partial x_i$, continuity of electric
potential at the domain wall $\mu = \1{\varphi} = \2{\varphi}$,
and continuity of electric displacement at the domain wall
$\DifrOp{P_i + \varepsilon_\mathrm{b}E_i} n_i = \sigma_f$. Combining the
above expressions, the formula for the local force density of external
sources on the domain wall can be written in a form
\begin{equation}
    \label{eq:18:pIs}
    f_k = \DifrOp{
                \Phi(P) -
                \frac 12 \varepsilon_\mathrm{b} E_i E_i -
                E_i P_i
            } n_k +
            \DifrOp{
                E_k \left(P_i + \varepsilon_\mathrm{b}E_i\right)
            } n_i.
\end{equation}

Considering the continuity of tangential components of electric
field at the domain wall $\DifrOp{E_{t,i}} = \DifrOp{E_i -
(E_kn_k) n_i} = 0$ and using the algebraic identity
\begin{equation}
    \label{eq:19:AlgebrId}
    \DifrOp{fg} = \widehat{f}\DifrOp{g} + \DifrOp{f}\widehat{g},
\end{equation}
where $\widehat{f} = \left(\1{f} + \2{f}\right)/2$ is the average of
bulk quantity at the opposite sides of the domain wall, the
general formula for the local force density of external sources on the
domain wall can be further simplified to the form:
\begin{equation}
    \label{app_force_simplified}
    f_k =
    \DifrOp{\Phi(P)} n_k
        - \widehat{E}_i \DifrOp{P_i} n_k +
       \widehat{E}_k \sigma_\mathrm{f}.
\end{equation}


\begin{thebibliography}{10}%
\makeatletter
\providecommand \@ifxundefined [1]{%
 \ifx #1\undefined \expandafter \@firstoftwo
 \else \expandafter \@secondoftwo
\fi
}%
\providecommand \@ifnum [1]{%
 \ifnum #1\expandafter \@firstoftwo
 \else \expandafter \@secondoftwo
\fi
}%
\providecommand \enquote [1]{``#1''}%
\providecommand \bibnamefont  [1]{#1}%
\providecommand \bibfnamefont [1]{#1}%
\providecommand \citenamefont [1]{#1}%
\providecommand\href[0]{\@sanitize\@href}%
\providecommand\@href[1]{\endgroup\@@startlink{#1}\endgroup\@@href}%
\providecommand\@@href[1]{#1\@@endlink}%
\providecommand \@sanitize [0]{\begingroup\catcode`\&12\catcode`\#12\relax}%
\@ifxundefined \pdfoutput {\@firstoftwo}{%
 \@ifnum{\z@=\pdfoutput}{\@firstoftwo}{\@secondoftwo}%
}{%
 \providecommand\@@startlink[1]{\leavevmode}%
 \providecommand\@@endlink[0]{}%
}{%
 \providecommand\@@startlink[1]{%
  \leavevmode
  \pdfstartlink
   attr{/Border[0 0 1 ]/H/I/C[0 1 1]}%
   user{/Subtype/Link/A<</Type/Action/S/URI/URI(#1)>>}%
  \relax
 }%
 \providecommand\@@endlink[0]{\pdfendlink}%
}%
\providecommand \url  [0]{\begingroup\@sanitize \@url }%
\providecommand \@url [1]{\endgroup\@href {#1}{\urlprefix}}%
\providecommand \urlprefix [0]{URL }%
\providecommand \Eprint[0]{\href }%
\@ifxundefined \urlstyle {%
  \providecommand \doi [1]{doi:\discretionary{}{}{}#1}%
}{%
  \providecommand \doi [0]{doi:\discretionary{}{}{}\begingroup
  \urlstyle{rm}\Url }%
}%
\providecommand \doibase [0]{http://dx.doi.org/}%
\providecommand \Doi[1]{\href{\doibase#1}}%
\providecommand \bibAnnote [3]{%
  \BibitemShut{#1}%
  \begin{quotation}\noindent
    \textsc{Key:}\ #2\\\textsc{Annotation:}\ #3%
  \end{quotation}%
}%
\providecommand \bibAnnoteFile [2]{%
  \IfFileExists{#2}{\bibAnnote {#1} {#2} {\input{#2}}}{}%
}%
\providecommand \typeout [0]{\immediate \write \m@ne }%
\providecommand \selectlanguage [0]{\@gobble}%
\providecommand \bibinfo [0]{\@secondoftwo}%
\providecommand \bibfield [0]{\@secondoftwo}%
\providecommand \translation [1]{[#1]}%
\providecommand \BibitemOpen[0]{}%
\providecommand \bibitemStop [0]{}%
\providecommand \bibitemNoStop [0]{.\EOS\space}%
\providecommand \EOS [0]{\spacefactor3000\relax}%
\providecommand \BibitemShut [1]{\csname bibitem#1\endcsname}%
%</preamble>
\bibitem{Tagantsev_book}%
  \BibitemOpen
  \bibfield{author}{%
  \bibinfo {author} {\bibfnamefont{A.~A.}\ \bibnamefont{Tagantsev}}, \bibinfo
  {author} {\bibfnamefont{L.~E.}\ \bibnamefont{Cross}},\ and\ \bibinfo {author}
  {\bibfnamefont{J.}~\bibnamefont{Fousek}},\ }%
  \emph{\bibinfo {title} {Domains in Ferroic Crystals and Thin Films}}\
  (\bibinfo {publisher} {Springer},\ \bibinfo {address} {New York},\ \bibinfo
  {year} {2010})\ Chap.\ \bibinfo {chapter} {5.2, 6.1}%
  \bibAnnoteFile{NoStop}{Tagantsev_book}%
\bibitem{Fesenko73}%
  \BibitemOpen
  \bibfield{author}{%
  \bibinfo {author} {\bibfnamefont{E.~G.}\ \bibnamefont{Fesenko}}, \bibinfo
  {author} {\bibfnamefont{V.~G.}\ \bibnamefont{Gavrilyatchenko}}, \bibinfo
  {author} {\bibfnamefont{M.~A.}\ \bibnamefont{Martinenko}}, \bibinfo {author}
  {\bibfnamefont{A.~F.}\ \bibnamefont{Semenchov}},\ and\ \bibinfo {author}
  {\bibfnamefont{I.~P.}\ \bibnamefont{Lapin}},\ }%
  \bibfield{journal}{%
  \bibinfo {journal} {Ferroelectrics}\ }%
  \textbf{\bibinfo {volume} {6}},\ \bibinfo {pages} {61} (\bibinfo {year}
  {1973})%
  \bibAnnoteFile{NoStop}{Fesenko73}%
\bibitem{Fesenko85}%
  \BibitemOpen
  \bibfield{author}{%
  \bibinfo {author} {\bibfnamefont{E.~G.}\ \bibnamefont{Fesenko}}, \bibinfo
  {author} {\bibfnamefont{V.~G.}\ \bibnamefont{Gavrilyatchenko}}, \bibinfo
  {author} {\bibfnamefont{A.~F.}\ \bibnamefont{Semenchov}},\ and\ \bibinfo
  {author} {\bibfnamefont{S.~M.}\ \bibnamefont{Yufatova}},\ }%
  \bibfield{journal}{%
  \bibinfo {journal} {Ferroelectrics}\ }%
  \textbf{\bibinfo {volume} {63}},\ \bibinfo {pages} {289} (\bibinfo {year}
  {1985})%
  \bibAnnoteFile{NoStop}{Fesenko85}%
\bibitem{Surowiak}%
  \BibitemOpen
  \bibfield{author}{%
  \bibinfo {author} {\bibfnamefont{Z.}~\bibnamefont{Surowiak}}, \bibinfo
  {author} {\bibfnamefont{J.}~\bibnamefont{Dec}}, \bibinfo {author}
  {\bibfnamefont{R.}~\bibnamefont{Skulski}}, \bibinfo {author}
  {\bibfnamefont{E.~G.}\ \bibnamefont{Fesenko}}, \bibinfo {author}
  {\bibfnamefont{V.~G.}\ \bibnamefont{Gavrilyatchenko}},\ and\ \bibinfo
  {author} {\bibfnamefont{A.~F.}\ \bibnamefont{Semenchov}},\ }%
  \bibfield{journal}{%
  \bibinfo {journal} {Ferroelectrics}\ }%
  \textbf{\bibinfo {volume} {20}},\ \bibinfo {pages} {277} (\bibinfo {year}
  {1978})%
  \bibAnnoteFile{NoStop}{Surowiak}%
\bibitem{Jia}%
  \BibitemOpen
  \bibfield{author}{%
  \bibinfo {author} {\bibfnamefont{C.-L.}\ \bibnamefont{Jia}}, \bibinfo
  {author} {\bibfnamefont{S.-B.}\ \bibnamefont{Mi}}, \bibinfo {author}
  {\bibfnamefont{K.}~\bibnamefont{Urban}}, \bibinfo {author}
  {\bibfnamefont{I.}~\bibnamefont{Vrejoiu}}, \bibinfo {author}
  {\bibfnamefont{M.}~\bibnamefont{Alexe}},\ and\ \bibinfo {author}
  {\bibfnamefont{D.}~\bibnamefont{Hesse}},\ }%
  \bibfield{journal}{%
  \bibinfo {journal} {Nature Materials}\ }%
  \textbf{\bibinfo {volume} {7}},\ \bibinfo {pages} {57} (\bibinfo {year}
  {2008})%
  \bibAnnoteFile{NoStop}{Jia}%
\bibitem{Shur93}%
  \BibitemOpen
  \bibfield{author}{%
  \bibinfo {author} {\bibfnamefont{V.~Y.}\ \bibnamefont{Shur}}, \bibinfo
  {author} {\bibfnamefont{E.~L.}\ \bibnamefont{Rumyantsev}},\ and\ \bibinfo
  {author} {\bibfnamefont{A.~L.}\ \bibnamefont{Subbotin}},\ }%
  \bibfield{journal}{%
  \bibinfo {journal} {Ferroelectrics}\ }%
  \textbf{\bibinfo {volume} {140}},\ \bibinfo {pages} {305} (\bibinfo {year}
  {1993})%
  \bibAnnoteFile{NoStop}{Shur93}%
\bibitem{Shur}%
  \BibitemOpen
  \bibfield{author}{%
  \bibinfo {author} {\bibfnamefont{V.}~\bibnamefont{Shur}}\ and\ \bibinfo
  {author} {\bibfnamefont{E.}~\bibnamefont{Rumyantsev}},\ }%
  \bibfield{journal}{%
  \bibinfo {journal} {Journal of the Korean Physical Society}\ }%
  \textbf{\bibinfo {volume} {32}},\ \bibinfo {pages} {S727} (\bibinfo {year}
  {1998})%
  \bibAnnoteFile{NoStop}{Shur}%
\bibitem{kugel}%
  \BibitemOpen
  \bibfield{author}{%
  \bibinfo {author} {\bibfnamefont{V.~D.}\ \bibnamefont{Kugel}}\ and\ \bibinfo
  {author} {\bibfnamefont{G.}~\bibnamefont{Rosenman}},\ }%
  \bibfield{journal}{%
  \Doi{10.1063/1.109191}{\bibinfo {journal} {Applied Physics Letters}}\ }%
  \textbf{\bibinfo {volume} {62}},\ \bibinfo {pages} {2902} (\bibinfo {year}
  {1993})%
  \bibAnnoteFile{NoStop}{kugel}%
\bibitem{nakamura}%
  \BibitemOpen
  \bibfield{author}{%
  \bibinfo {author} {\bibfnamefont{K.}~\bibnamefont{Nakamura}}, \bibinfo
  {author} {\bibfnamefont{H.}~\bibnamefont{Ando}},\ and\ \bibinfo {author}
  {\bibfnamefont{H.}~\bibnamefont{Shimizu}},\ }%
  \bibfield{journal}{%
  \Doi{10.1063/1.97838}{\bibinfo {journal} {Applied Physics Letters}}\ }%
  \textbf{\bibinfo {volume} {50}},\ \bibinfo {pages} {1413} (\bibinfo {year}
  {1987})%
  \bibAnnoteFile{NoStop}{nakamura}%
\bibitem{Vul}%
  \BibitemOpen
  \bibfield{author}{%
  \bibinfo {author} {\bibfnamefont{B.~M.}\ \bibnamefont{Vul}}, \bibinfo
  {author} {\bibfnamefont{G.~M.}\ \bibnamefont{Guro}},\ and\ \bibinfo {author}
  {\bibfnamefont{I.~I.}\ \bibnamefont{Ivanchik}},\ }%
  \bibfield{journal}{%
  \bibinfo {journal} {Ferroelectrics}\ }%
  \textbf{\bibinfo {volume} {6}},\ \bibinfo {pages} {29} (\bibinfo {year}
  {1973})%
  \bibAnnoteFile{NoStop}{Vul}%
\bibitem{Ivanchik}%
  \BibitemOpen
  \bibfield{author}{%
  \bibinfo {author} {\bibfnamefont{I.~I.}\ \bibnamefont{Ivanchik}},\ }%
  \bibfield{journal}{%
  \bibinfo {journal} {Ferroelectrics}\ }%
  \textbf{\bibinfo {volume} {145}},\ \bibinfo {pages} {149} (\bibinfo {year}
  {1993})%
  \bibAnnoteFile{NoStop}{Ivanchik}%
\bibitem{Morozovska}%
  \BibitemOpen
  \bibfield{author}{%
  \bibinfo {author} {\bibfnamefont{E.~A.}\ \bibnamefont{Eliseev}}, \bibinfo
  {author} {\bibfnamefont{A.~N.}\ \bibnamefont{Morozovska}}, \bibinfo {author}
  {\bibfnamefont{G.~S.}\ \bibnamefont{Svechnikov}}, \bibinfo {author}
  {\bibfnamefont{V.}~\bibnamefont{Gopalan}},\ and\ \bibinfo {author}
  {\bibfnamefont{V.~Y.}\ \bibnamefont{Shur}},\ }%
  \bibfield{journal}{%
  \bibinfo {journal} {Physical Review B}\ }%
  \textbf{\bibinfo {volume} {83}},\ \bibinfo {pages} {235313}  (\bibinfo {year} {2011})%
  \bibAnnoteFile{NoStop}{Morozovska}%
\bibitem{Landauer}%
  \BibitemOpen
  \bibfield{author}{%
  \bibinfo {author} {\bibfnamefont{R.}~\bibnamefont{Landauer}},\ }%
  \bibfield{journal}{%
  \bibinfo {journal} {J. Appl. Phys.}\ }%
  \textbf{\bibinfo {volume} {28}},\ \bibinfo {pages} {227} (\bibinfo {year}
  {1957})%
  \bibAnnoteFile{NoStop}{Landauer}%
\bibitem{Mokry}%
  \BibitemOpen
  \bibfield{author}{%
  \bibinfo {author} {\bibfnamefont{P.}~\bibnamefont{Mokr\'{y}}}, \bibinfo
  {author} {\bibfnamefont{A.~K.}\bibnamefont{Tagantsev}},\ and\ \bibinfo {author}
  {\bibfnamefont{J.}~\bibnamefont{Fousek}},\ }%
  \bibfield{journal}{%
  \bibinfo {journal} {Physical Review B}\ }%
    \textbf{\bibinfo {volume} {75}},\ \bibinfo {pages} {094110} (\bibinfo {year}
  {2007})%
  \bibAnnoteFile{NoStop}{Mokry}%
\bibitem{Landau}%
  \BibitemOpen
  \bibfield{author}{%
  \bibinfo {author} {\bibfnamefont{L.~D.}\ \bibnamefont{Landau}}\ and\ \bibinfo
  {author} {\bibfnamefont{E.~M.}\ \bibnamefont{Lifshits}},\ }%
  \emph{\bibinfo {title} {Electrodynamics of Continuous Media}}\ (\bibinfo
  {publisher} {Pergamon Press},\ \bibinfo {address} {Oxford},\ \bibinfo {year}
  {1984})%
  \bibAnnoteFile{NoStop}{Landau}%
\bibitem{Landau_comment}%
  \BibitemOpen
  \bibinfo {note} {The direct application of the formula for the generalized
  stress tensor, actually given in the book by Landau and Lifshitz\cite{Landau}
  without derivation, leads to the force acting on the domain wall expressed as
  follows: $ f_k = [[\Phi(P)]]-\hat{E}_i[[P_i]]n_k+\hat{E}_k\sigma_\mathrm{f}
  +[[P_kE_i-P_iE_k]]n_i$. In comparison to the result, obtained using the
  method of virtual displacement, this force contains an additional (in general
  non-zero) term $[[P_kE_i-P_iE_k]]n_i$. For the case of the pressure this term
  is equal to zero and the result obtained with the method of virtual
  displacement is the same as that obtained using the stress tensor from the
  book by Landau and Lifshitz.}%
  \bibAnnoteFile{Stop}{Landau_comment}%
\bibitem{Gureev}%
  \BibitemOpen
  \bibfield{author}{%
  \bibinfo {author} {\bibfnamefont{M.~Y.}\ \bibnamefont{Gureev}}, \bibinfo {author}
  {\bibfnamefont{A.~K.}\ \bibnamefont{Tagantsev}},\ and\ \bibinfo {author}
  {\bibfnamefont{N.}~\bibnamefont{Setter}},\ }%
  \bibfield{journal}{%
  \bibinfo {journal} {Physical Review B}\ }%
  \textbf{\bibinfo {volume} {83}},\ \bibinfo {pages} {184104}  (\bibinfo {year} {2011})%
  \bibAnnoteFile{NoStop}{Gureev}%
\bibitem{Guro68}%
  \BibitemOpen
  \bibfield{author}{%
  \bibinfo {author} {\bibfnamefont{G.~M.}\ \bibnamefont{Guro}}, \bibinfo
  {author} {\bibfnamefont{I.~I.}\ \bibnamefont{Ivanchik}},\ and\ \bibinfo
  {author} {\bibfnamefont{N.~F.}\ \bibnamefont{Kovtonyuk}},\ }%
  \bibfield{journal}{%
  \bibinfo {journal} {Sov. Phys. - Solid State}\ }%
  \textbf{\bibinfo {volume} {10}},\ \bibinfo {pages} {100} (\bibinfo {year}
  {1968})%
  \bibAnnoteFile{NoStop}{Guro68}%
\bibitem{minimal_period}%
  \BibitemOpen
  \bibfield{author}{%
  \bibinfo {author} {\bibfnamefont{M.~Y.}\ \bibnamefont{Gureev}}, \bibinfo {author}
  {\bibfnamefont{A.~K.}\ \bibnamefont{Tagantsev}},\ and\ \bibinfo {author}
  {\bibfnamefont{N.}~\bibnamefont{Setter}},\ }%
  \bibinfo {howpublished} {Unpublished}%
  \bibAnnoteFile{NoStop}{minimal_period}%
\bibitem{Merz}%
  \BibitemOpen
  \bibfield{author}{%
  \bibinfo {author} {\bibfnamefont{W.~J.}\ \bibnamefont{Merz}},\ }%
  \bibfield{journal}{%
  \bibinfo {journal} {Phys. Rev.}\ }%
  \textbf{\bibinfo {volume} {95}},\ \bibinfo {pages} {690} (\bibinfo {year}
  {1954})%
  \bibAnnoteFile{NoStop}{Merz}%
\end{thebibliography}
\end{document}